\documentclass[%
 reprint,
superscriptaddress,
notitlepage,
showpacs,preprintnumbers,
 amsmath,amssymb,
 aps,
floatfix,
]{revtex4-1}

\usepackage{amssymb, amsmath} % introduce ams math environments
\usepackage{graphicx}% Include figure files
\usepackage{dcolumn}% Align table columns on decimal point
\usepackage{bm}% bold math
\usepackage{color}
\usepackage[caption=false]{subfig}
\usepackage{natbib}
\usepackage{hyperref}
\usepackage[capitalize]{cleveref}
\usepackage{textgreek} % type greek letters in text mode with \textbeta
\usepackage{siunitx}
\usepackage[version=3]{mhchem}

\newcommand*{\cczpi}{CC0$\pi$}
\newcommand*{\ccopi}{CC1$\pi^+$}

\newcommand*{\numu}{\ce{$\nu$_{$\mu$}}}

\newcommand*{\pip}{\ce{$\pi$+}}

\newcommand*{\pd}{P\O{}D}

\newcommand*{\minerva}{MINER\textnu{}A}
\newcommand*{\miniboone}{MiniBooNE}

\begin{document}

%\linenumbers

\title{First measurement of the $\nu_{\mu}$ charged-current cross section on a water target without pions in the final state }% Force line breaks with \\

%%%%%%%%%%%%%%%%%%%%%%%%%%%%%%%%%%%%%%%%%%%%%%%%%%%%%%%%%%%%%%
% T2K author list generated on Sat, 13 Jan 2018 04:40:19 +0900
% setting: extra = 0 revtex = 1 ptep = 0 simple = 0 xml = 0 yearrule = 1 shiftrule = 1
%         author list from archive (starting 2017/05/15 until 2017/10/31)
%         exemption(s) granted to: tianlu
% Number of authors = 305
%%%%%%%%%%%%%%%%%%%%%%%%%%%%%%%%%%%%%%%%%%%%%%%%%%%%%%%%%%%%%%

\newcommand{\INSTEE}{\affiliation{University of Bern, Albert Einstein Center for Fundamental Physics, Laboratory for High Energy Physics (LHEP), Bern, Switzerland}}
\newcommand{\INSTFE}{\affiliation{Boston University, Department of Physics, Boston, Massachusetts, U.S.A.}}
\newcommand{\INSTD}{\affiliation{University of British Columbia, Department of Physics and Astronomy, Vancouver, British Columbia, Canada}}
\newcommand{\INSTGA}{\affiliation{University of California, Irvine, Department of Physics and Astronomy, Irvine, California, U.S.A.}}
\newcommand{\INSTI}{\affiliation{IRFU, CEA Saclay, Gif-sur-Yvette, France}}
\newcommand{\INSTGB}{\affiliation{University of Colorado at Boulder, Department of Physics, Boulder, Colorado, U.S.A.}}
\newcommand{\INSTFG}{\affiliation{Colorado State University, Department of Physics, Fort Collins, Colorado, U.S.A.}}
\newcommand{\INSTFH}{\affiliation{Duke University, Department of Physics, Durham, North Carolina, U.S.A.}}
\newcommand{\INSTBA}{\affiliation{Ecole Polytechnique, IN2P3-CNRS, Laboratoire Leprince-Ringuet, Palaiseau, France }}
\newcommand{\INSTEG}{\affiliation{University of Geneva, Section de Physique, DPNC, Geneva, Switzerland}}
\newcommand{\INSTDG}{\affiliation{H. Niewodniczanski Institute of Nuclear Physics PAN, Cracow, Poland}}
\newcommand{\INSTCB}{\affiliation{High Energy Accelerator Research Organization (KEK), Tsukuba, Ibaraki, Japan}}
\newcommand{\INSTED}{\affiliation{Institut de Fisica d'Altes Energies (IFAE), The Barcelona Institute of Science and Technology, Campus UAB, Bellaterra (Barcelona) Spain}}
\newcommand{\INSTEC}{\affiliation{IFIC (CSIC \& University of Valencia), Valencia, Spain}}
\newcommand{\INSTEI}{\affiliation{Imperial College London, Department of Physics, London, United Kingdom}}
\newcommand{\INSTGF}{\affiliation{INFN Sezione di Bari and Universit\`a e Politecnico di Bari, Dipartimento Interuniversitario di Fisica, Bari, Italy}}
\newcommand{\INSTBE}{\affiliation{INFN Sezione di Napoli and Universit\`a di Napoli, Dipartimento di Fisica, Napoli, Italy}}
\newcommand{\INSTBF}{\affiliation{INFN Sezione di Padova and Universit\`a di Padova, Dipartimento di Fisica, Padova, Italy}}
\newcommand{\INSTBD}{\affiliation{INFN Sezione di Roma and Universit\`a di Roma ``La Sapienza'', Roma, Italy}}
\newcommand{\INSTEB}{\affiliation{Institute for Nuclear Research of the Russian Academy of Sciences, Moscow, Russia}}
\newcommand{\INSTHA}{\affiliation{Kavli Institute for the Physics and Mathematics of the Universe (WPI), The University of Tokyo Institutes for Advanced Study, University of Tokyo, Kashiwa, Chiba, Japan}}
\newcommand{\INSTCC}{\affiliation{Kobe University, Kobe, Japan}}
\newcommand{\INSTCD}{\affiliation{Kyoto University, Department of Physics, Kyoto, Japan}}
\newcommand{\INSTEJ}{\affiliation{Lancaster University, Physics Department, Lancaster, United Kingdom}}
\newcommand{\INSTFC}{\affiliation{University of Liverpool, Department of Physics, Liverpool, United Kingdom}}
\newcommand{\INSTFI}{\affiliation{Louisiana State University, Department of Physics and Astronomy, Baton Rouge, Louisiana, U.S.A.}}
\newcommand{\INSTHB}{\affiliation{Michigan State University, Department of Physics and Astronomy,  East Lansing, Michigan, U.S.A.}}
\newcommand{\INSTCE}{\affiliation{Miyagi University of Education, Department of Physics, Sendai, Japan}}
\newcommand{\INSTDF}{\affiliation{National Centre for Nuclear Research, Warsaw, Poland}}
\newcommand{\INSTFJ}{\affiliation{State University of New York at Stony Brook, Department of Physics and Astronomy, Stony Brook, New York, U.S.A.}}
\newcommand{\INSTGJ}{\affiliation{Okayama University, Department of Physics, Okayama, Japan}}
\newcommand{\INSTCF}{\affiliation{Osaka City University, Department of Physics, Osaka, Japan}}
\newcommand{\INSTGG}{\affiliation{Oxford University, Department of Physics, Oxford, United Kingdom}}
\newcommand{\INSTBB}{\affiliation{UPMC, Universit\'e Paris Diderot, CNRS/IN2P3, Laboratoire de Physique Nucl\'eaire et de Hautes Energies (LPNHE), Paris, France}}
\newcommand{\INSTGC}{\affiliation{University of Pittsburgh, Department of Physics and Astronomy, Pittsburgh, Pennsylvania, U.S.A.}}
\newcommand{\INSTFA}{\affiliation{Queen Mary University of London, School of Physics and Astronomy, London, United Kingdom}}
\newcommand{\INSTE}{\affiliation{University of Regina, Department of Physics, Regina, Saskatchewan, Canada}}
\newcommand{\INSTGD}{\affiliation{University of Rochester, Department of Physics and Astronomy, Rochester, New York, U.S.A.}}
\newcommand{\INSTHC}{\affiliation{Royal Holloway University of London, Department of Physics, Egham, Surrey, United Kingdom}}
\newcommand{\INSTBC}{\affiliation{RWTH Aachen University, III. Physikalisches Institut, Aachen, Germany}}
\newcommand{\INSTFB}{\affiliation{University of Sheffield, Department of Physics and Astronomy, Sheffield, United Kingdom}}
\newcommand{\INSTDI}{\affiliation{University of Silesia, Institute of Physics, Katowice, Poland}}
\newcommand{\INSTEH}{\affiliation{STFC, Rutherford Appleton Laboratory, Harwell Oxford,  and  Daresbury Laboratory, Warrington, United Kingdom}}
\newcommand{\INSTCH}{\affiliation{University of Tokyo, Department of Physics, Tokyo, Japan}}
\newcommand{\INSTBJ}{\affiliation{University of Tokyo, Institute for Cosmic Ray Research, Kamioka Observatory, Kamioka, Japan}}
\newcommand{\INSTCG}{\affiliation{University of Tokyo, Institute for Cosmic Ray Research, Research Center for Cosmic Neutrinos, Kashiwa, Japan}}
\newcommand{\INSTGI}{\affiliation{Tokyo Metropolitan University, Department of Physics, Tokyo, Japan}}
\newcommand{\INSTF}{\affiliation{University of Toronto, Department of Physics, Toronto, Ontario, Canada}}
\newcommand{\INSTB}{\affiliation{TRIUMF, Vancouver, British Columbia, Canada}}
\newcommand{\INSTG}{\affiliation{University of Victoria, Department of Physics and Astronomy, Victoria, British Columbia, Canada}}
\newcommand{\INSTDJ}{\affiliation{University of Warsaw, Faculty of Physics, Warsaw, Poland}}
\newcommand{\INSTDH}{\affiliation{Warsaw University of Technology, Institute of Radioelectronics, Warsaw, Poland}}
\newcommand{\INSTFD}{\affiliation{University of Warwick, Department of Physics, Coventry, United Kingdom}}
\newcommand{\INSTGH}{\affiliation{University of Winnipeg, Department of Physics, Winnipeg, Manitoba, Canada}}
\newcommand{\INSTEA}{\affiliation{Wroclaw University, Faculty of Physics and Astronomy, Wroclaw, Poland}}
\newcommand{\INSTHE}{\affiliation{Yokohama National University, Faculty of Engineering, Yokohama, Japan}}
\newcommand{\INSTH}{\affiliation{York University, Department of Physics and Astronomy, Toronto, Ontario, Canada}}

\INSTEE
\INSTFE
\INSTD
\INSTGA
\INSTI
\INSTGB
\INSTFG
\INSTFH
\INSTBA
\INSTEG
\INSTDG
\INSTCB
\INSTED
\INSTEC
\INSTEI
\INSTGF
\INSTBE
\INSTBF
\INSTBD
\INSTEB
\INSTHA
\INSTCC
\INSTCD
\INSTEJ
\INSTFC
\INSTFI
\INSTHB
\INSTCE
\INSTDF
\INSTFJ
\INSTGJ
\INSTCF
\INSTGG
\INSTBB
\INSTGC
\INSTFA
\INSTE
\INSTGD
\INSTHC
\INSTBC
\INSTFB
\INSTDI
\INSTEH
\INSTCH
\INSTBJ
\INSTCG
\INSTGI
\INSTF
\INSTB
\INSTG
\INSTDJ
\INSTDH
\INSTFD
\INSTGH
\INSTEA
\INSTHE
\INSTH

\author{K.\,Abe}\INSTBJ
\author{J.\,Amey}\INSTEI
\author{C.\,Andreopoulos}\INSTEH\INSTFC
\author{L.\,Anthony}\INSTFC
\author{M.\,Antonova}\INSTEC
\author{S.\,Aoki}\INSTCC
\author{A.\,Ariga}\INSTEE
\author{Y.\,Ashida}\INSTCD
\author{S.\,Ban}\INSTCD
\author{M.\,Barbi}\INSTE
% ...

\collaboration{The T2K Collaboration}\noaffiliation

%\collaboration{The T2K Collaboration}\noaffiliation

\date{\today}      %  but any date may be explicitly specified

\begin{abstract}

This paper reports the first differential measurement of the charged-current interaction cross section of $\nu_{\mu}$ on water with no pions in the final state. This flux-averaged measurement has been made using the T2K experiment's off-axis near detector, and is reported in doubly-differential bins of muon momentum and angle.  The flux-averaged total cross section in a restricted region of phase space was  found to be $ \sigma= \SI[parse-numbers=false]{(0.95 \pm{} 0.08 (stat) \pm{} 0.06 (\mbox{det. syst.}) \pm{} 0.04(\mbox{model syst.}) \pm{} 0.08(flux)  )\times 10^{-38}}{\cm^2 \per n}$.

\end{abstract}

% check these!!!!
\maketitle

\section{\label{sec:intro}Introduction}
The T2K (Tokai-to-Kamioka) experiment\cite{t2knim} is a long-baseline neutrino oscillation experiment, with a beam originating at J-PARC (Japan Proton Accelerator Complex) which consists primarily of muon neutrinos. T2K has measured the disappearance of muon neutrinos~\cite{t2knumu} and the appearance of electron neutrinos~\cite{t2knue}, using the off-axis ND280 near detector on the J-PARC site, and the Super-Kamiokande detector~\cite{sknim}, located \SI{295}{\km} away.

At the energies of the T2K beamline, the main neutrino interaction process is charged-current quasi-elastic interactions ($\nu_{\mu} + n \rightarrow \mu^{-} + p$).  Because these neutrino interactions occur within nuclear targets and not on free nuclei, additional nuclear effects and final state interactions can modify the composition and kinematics of the particles that are observed to be exiting the interaction.  This paper focuses on a measurement of CCQE-like events (in bins of the muon angle and momentum) in which no pions (charged or neutral) are observed in the final state (\cczpi{}).

The active target regions of the ND280 near detector, as will be discussed in Section~\ref{sec:t2k}, are primarily composed of plastic scintillator, but the far detector is water-based.  While the near detector measurements on hydrocarbon targets can help to greatly constrain the flux and cross section uncertainties for the oscillation analyses, one of the dominant remaining uncertainties is due to potential differences between the oxygen and carbon cross sections that are not currently well-constrained by the ND280 detector~\cite{nueapp}.  The ND280 detector also contains water targets, and this paper presents a measurement of the $\nu_{\mu}$ \cczpi{}  interaction cross section on water.  This process is very important for T2K's neutrino oscillation measurements since this is the dominant reaction in the far detector.

While there are differential measurements of the charged-current quasi-elastic (CCQE)  cross sections on carbon or hydrocarbon~\cite{miniboone_ccqe,sciboone_ccqe, minerva_ccqe, t2k_ccqe, hadley_ccqe}, there are none on water.  The K2K experiment has published a measurement of the axial vector mass in neutrino-oxygen CCQE interactions~\cite{k2k_ma} using the SciFi detector, but not a differential cross section measurement.

\section{\label{sec:t2k}The T2K Experiment}
A high-intensity beam is produced at J-PARC at Tokai village, Ibaraki, Japan and directed to Super-Kamiokande. 
In order to enhance the sensitivity to T2K's primary physics goals, 
the appearance of electron neutrinos and the disappearance of muon 
neutrinos, the beam energy peaks at the oscillation 
maximum of \SI{0.6}{\GeV}, and this is achieved by adopting the off-axis 
method~\cite{t2knim}. 

\subsection{The T2K Beamline}
High-intensity \SI{30}{\GeV} protons from the J-PARC accelerator strike a
graphite target every 2.48 seconds and produce charged pions and
kaons.  These pions and kaons are focused to the forward direction by three horn
magnets~\cite{horn}  and decay in the 96-m long
decay volume.  Neutrinos (anti-neutrinos) are produced from decays of
positively (negatively) charged pions.  Horn magnets select pions of
either sign by flipping the current direction.

T2K started data taking in 2010 with a beam of primarily muon neutrinos until May 2014
and then started data taking with a primarily muon anti-neutrino beam until May 2016, and is continuing to take data. 
The analysis presented here uses data taken in the neutrino mode.

The neutrino flux is calculated based on the measurement of primary proton beam profiles and 
hadron production data, including 
the measurement of the pion and kaon yields~\cite{na61_yields} by the NA61/SHINE experiment~\cite{na61_det} at CERN. 
%The predicted flux at ND280 detectors including the systematic uncertainties is shown in Fig.\ref{}. 
The total absolute flux uncertainty is about 10\% at the peak energy, and details of the flux calculation are described in Ref.~\cite{t2k_flux}.

\subsection{The T2K ND280 Near Detector}
\begin{figure}
  \centering 
   \includegraphics[width=0.5\textwidth]{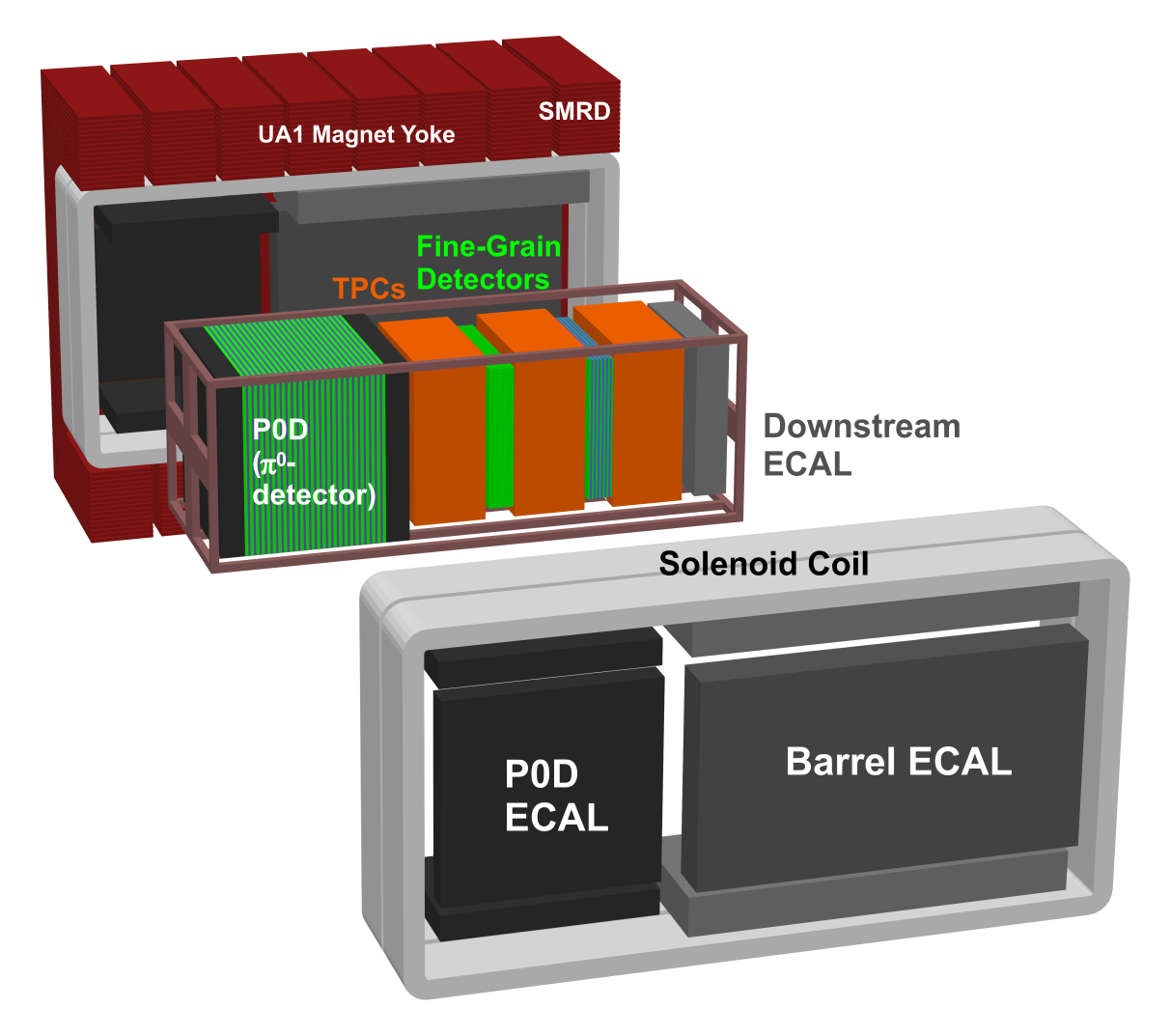}
    \caption{An exploded view of T2K's off-axis near detector.}
   \label{fig:nd280}
\end{figure}

\begin{figure}
  \centering 
   \includegraphics[width=0.5\textwidth]{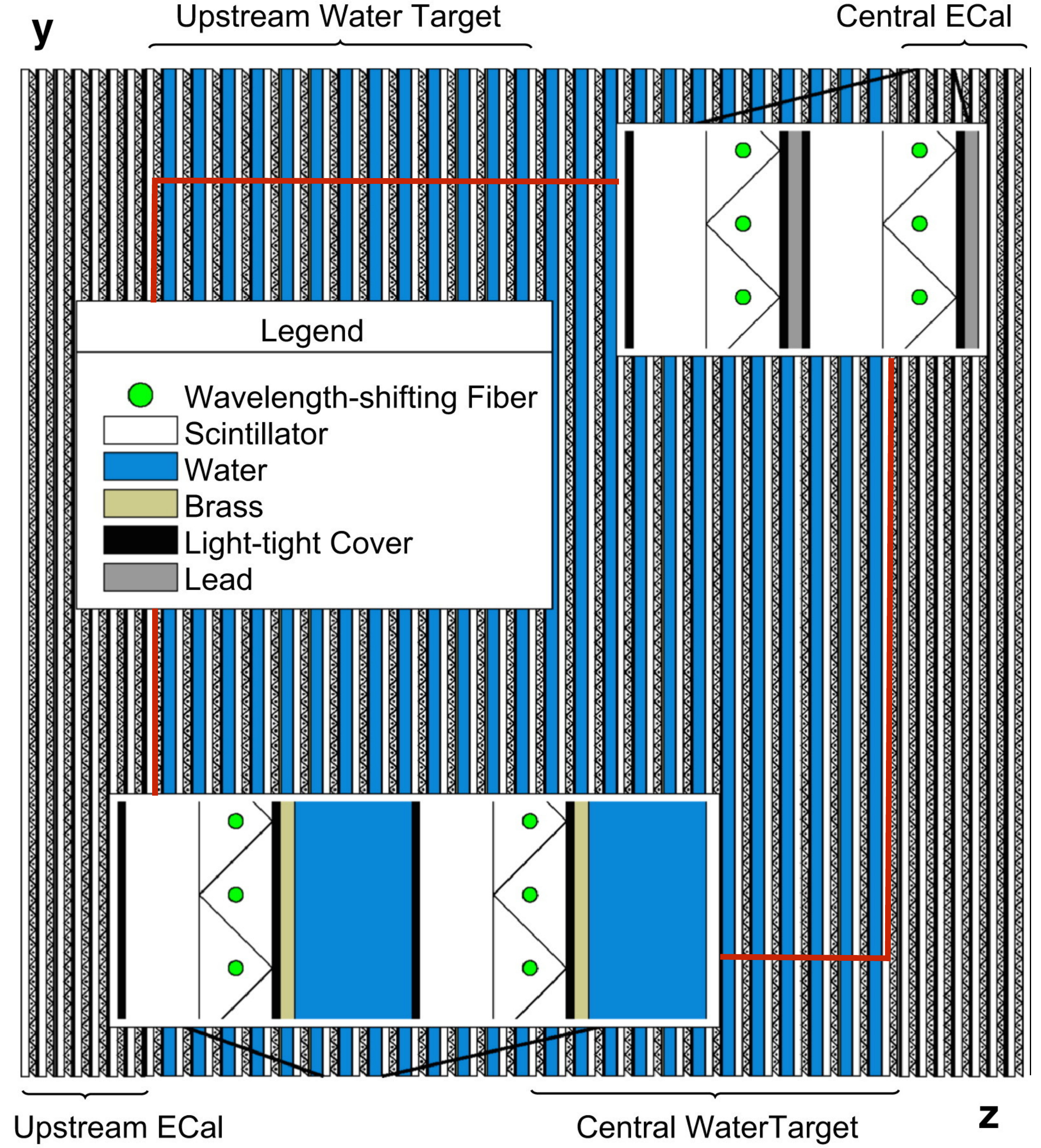}
    \caption{A schematic of the \pd{}, showing the structure of the layers.  The beam travels from the left to the right.  The red box outlines the approximate  fiducial volume used for this analysis.}
   \label{fig:p0d}
\end{figure}

This analysis looks for neutrino interactions in the T2K off-axis near detector, shown in Figure~\ref{fig:nd280}, specifically for events in the  pi-zero (\pd{}) subdetector.  This subdetector, described in more detail in~\cite{p0dnim} consists of alternating planes of scintillator bars, sandwiched between lead or brass radiator layers.  A more detailed schematic of the \pd{} is shown in Figure~\ref{fig:p0d}.  There are a total of 40 scintillator modules in the \pd{}, each one composed of a vertical layer of triangular scintillator bars and a layer of horizontal bars.  In the upstream and downstream portions of the \pd{} (the ``upstream ECal" and the ``central ECal"), the scintillator layers alternate with lead sheets.  The central 25 layers alternate the scintillator layers with brass sheets and bladders which can be filled with water or air.  This analysis will use data taken in both configurations. The total mass of water in the \pd{} in the fiducial volume is approximately 1902 kg. Downstream of the \pd{} is a tracker with three TPCs~\cite{tpcs} and two fine-grained scintillator detectors~\cite{fgds}.  The \pd{} and tracker are surrounded by electromagnetic calorimeters~\cite{ecal} inside a magnet with a 0.2 T field that is instrumented with muon detectors~\cite{smrd}.

\section{\label{sec:analysis}Analysis}

\subsection{Event Selection}

Hadrons produced within the nucleus are subject to final state
interactions (FSIs) that can reabsorb the hadron or alter their kinematics as they emerge from the nucleus. Pion absorption for example can make a charged-current resonant (CCRES)
interaction appear as a CCQE final state. This forces the selection to
be based on a topology classified by the number and type of outgoing
particles from the nucleus in order to be less model dependent. A single muon and zero pion topology is
called CCQE-like or \cczpi{}. 
%A measurement of the \cczpi{} cross section is more model independent than a measurement of CCQE.  %% CCQE can't be unambiguously measured.
The
event selection identifies \cczpi{}-enhanced samples by selecting
events where a single track was reconstructed in the \pd{}. Identical
selections are applied to both the water-in and water-out samples.

The result in this paper relies on a subset of the neutrino mode data
as broken down in \cref{tab:runs-pot} and totaling \num{5.52e20}
protons on the target (POT).  The selection starts by identifying beam spills where the spill
information and all ND280 subdetectors are known to have high-quality
data.  Each spill of the proton beam consists of several
clearly-separated bunches of protons.  The selection then selects
bunches containing tracks reconstructed in the TPCs associated with
vertices reconstructed within the \pd{} fiducial volume, which extends
from the middle of the scintillator of the first water layer to the middle of the scintillator in the last
water layer and \SI{25}{\cm} in from the edges in the xy-plane
(\cref{fig:p0d}).  In practice, these tracks are reconstructed from
segments found in the \pd{} and TPC subdetector reconstruction
processes. After identifying these bunches, the analysis identifies
the highest-momentum negatively-charged track as the muon
candidate. If no negatively charged track is found, the bunch is
removed from the selection.  The selection then applies a TPC
quality cut, removing tracks with no more than 18 nodes in the TPC
reconstruction.  The final cut for the selected  \cczpi{} sample requires that only a single track was reconstructed anywhere in the \pd{} in that
bunch.  The cross section reported here is restricted to bins in the region of muon kinematics where  $\cos \theta_\mu \geq 0$ and $p_\mu \leq \SI{5}{\GeV}$.  The total number of selected events in the water-in data sample is 12,777 (and based on simulations it is expected that approximately 3860 of these events are true interactions on water) and the total number in the water-out data (which has a larger POT exposure) sample is 13,370.  Due to the requirement of a negatively-charged track, the $\bar{\nu}_{\mu}$ contamination is this sample is very small and is estimated to be approximately 0.1\% for both samples.

\begin{table}
  \begin{ruledtabular}
  \begin{tabular}{lr}
    T2K Run & POT\\
    \hline
      2 Water & $4.29\times 10 ^{19}$  \\ 
    2 Air & $3.55\times 10 ^{19}$   \\ 
    3c Air & $1.35\times 10 ^{20}$   \\ 
    4 Water & $1.63\times 10 ^{20}$   \\ 
    4 Air & $1.76\times 10 ^{20}$   \\ 
    \hline
    Total & 5.52 $\times 10 ^{20}$  \\
  \end{tabular}
\end{ruledtabular}
  \caption[T2K runs and their associated POT]{T2K runs and their
    associated POT, filtered for spills where all ND280
    detectors were flagged with good data quality.}
  \label{tab:runs-pot}
\end{table}

\begin{figure}
  \centering
  \includegraphics[width=0.5\textwidth]{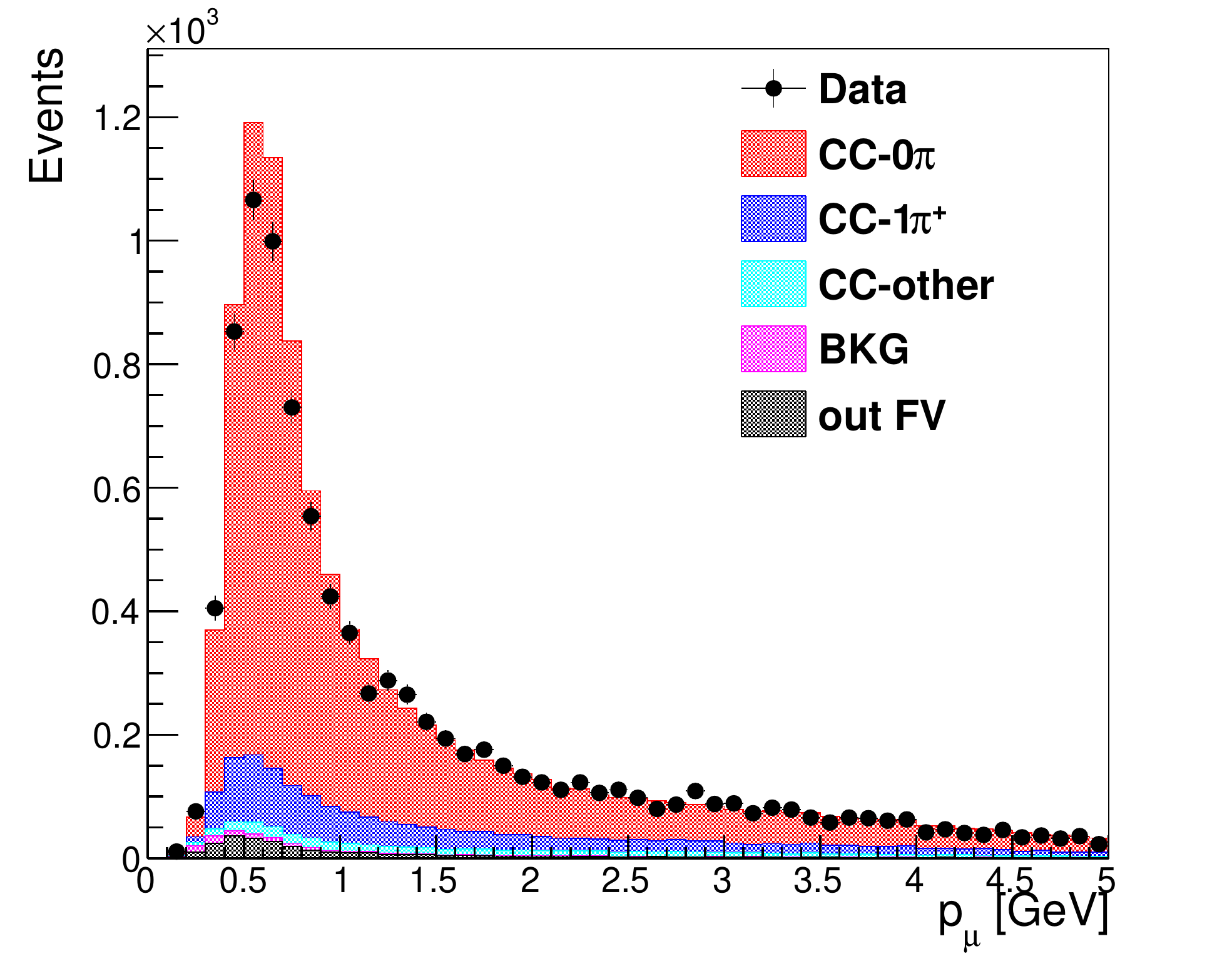}
  \caption[Data and MC distributions of the \cczpi{} selection]{Data
    and MC distributions of the water-in \cczpi{} signal
    selection. NEUT 5.3.2 MC has been normalized to data POT and is
    sorted into various truth topologies.}
\label{fig:cc0pisel}
\end{figure}

\Cref{fig:cc0pisel} shows distributions of events that pass the
selection for the water-in sample for data and a Monte Carlo detector simulation (MC)\@. The MC simulation chain primarily uses the NEUT neutrino event generator~\cite{neut} to provide the kinematics for particles emerging from neutrino interactions and a Geant4-based~\cite{geant4} \cite{geant42} package to simulate these particles moving through the detector (using Geant version 4.9.4).  The QGSP\_BERT model is used for hadronic interactions. 
The MC distribution is separated by interaction channel. A detailed review of neutrino interactions can be found in~\cite{reviewxs} for example.  The two largest sources of background are charged-current
interactions with a single outgoing \pip{} (\ccopi{}) and any other
charged-current interaction (CCOther) not categorized as \cczpi{} or
\ccopi{}. The \ccopi{} topology is due primarily to pion resonance
production and CCOther to deep inelastic scattering
(DIS)~\cite{reviewxs}. Additionally, neutral current (NC) backgrounds are classified as
``BKG'' and interactions occurring outside the \pd{} fiducial volume
are classified as ``out of FV'' or ``OOFV''. Distributions for the
water-out sample look similar.

\begin{figure}
  \centering
  \includegraphics[width=0.5\textwidth]{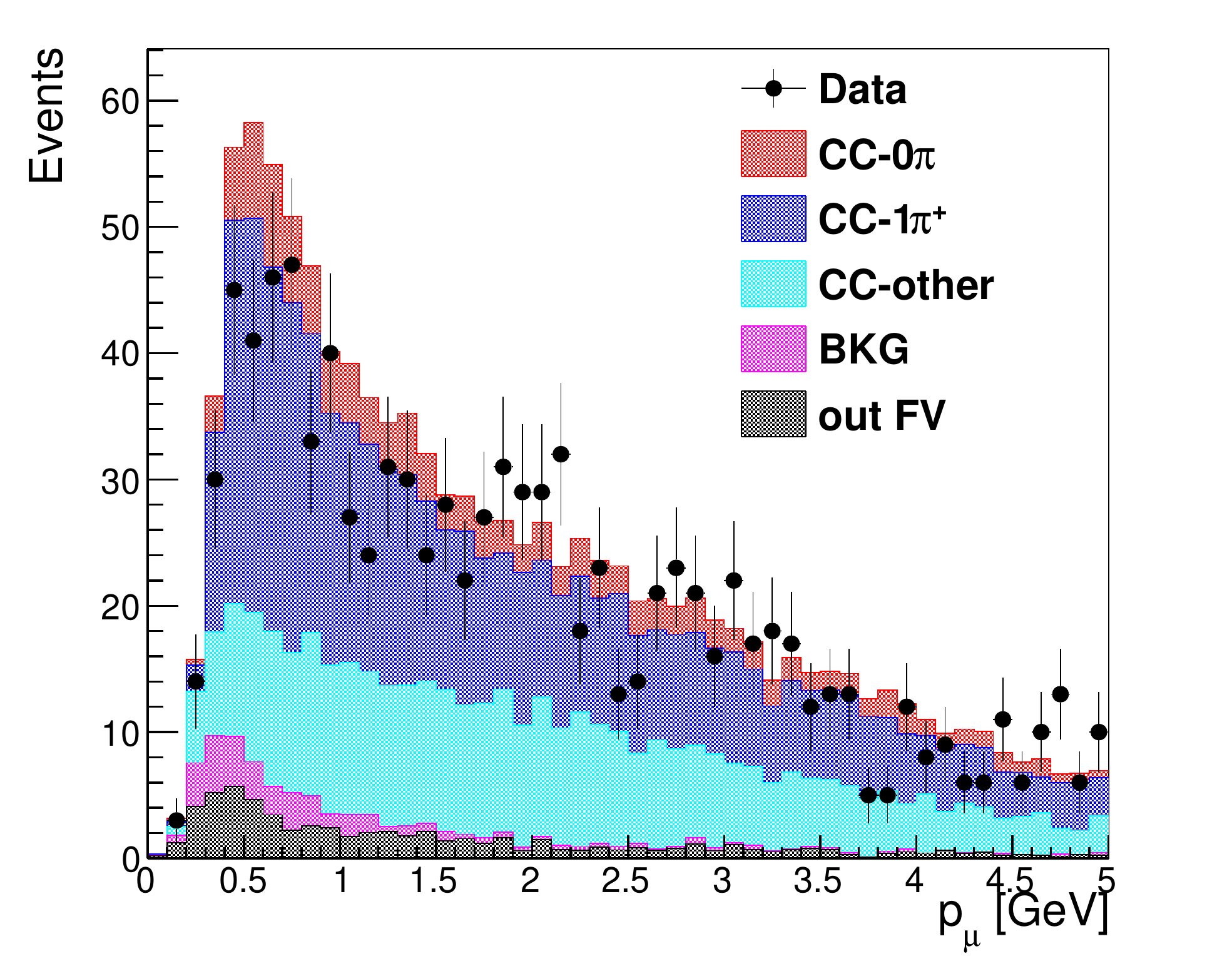}
  \caption[Data and MC distributions of the \ccopi{} selection]{Data
    and MC distributions of the water-in \ccopi{} sideband
    selection. NEUT 5.3.2 MC has been normalized to data POT and is
    sorted into various truth topologies.}
\label{fig:cc1pisel}
\end{figure}

\begin{figure}
  \centering
  \includegraphics[width=0.5\textwidth]{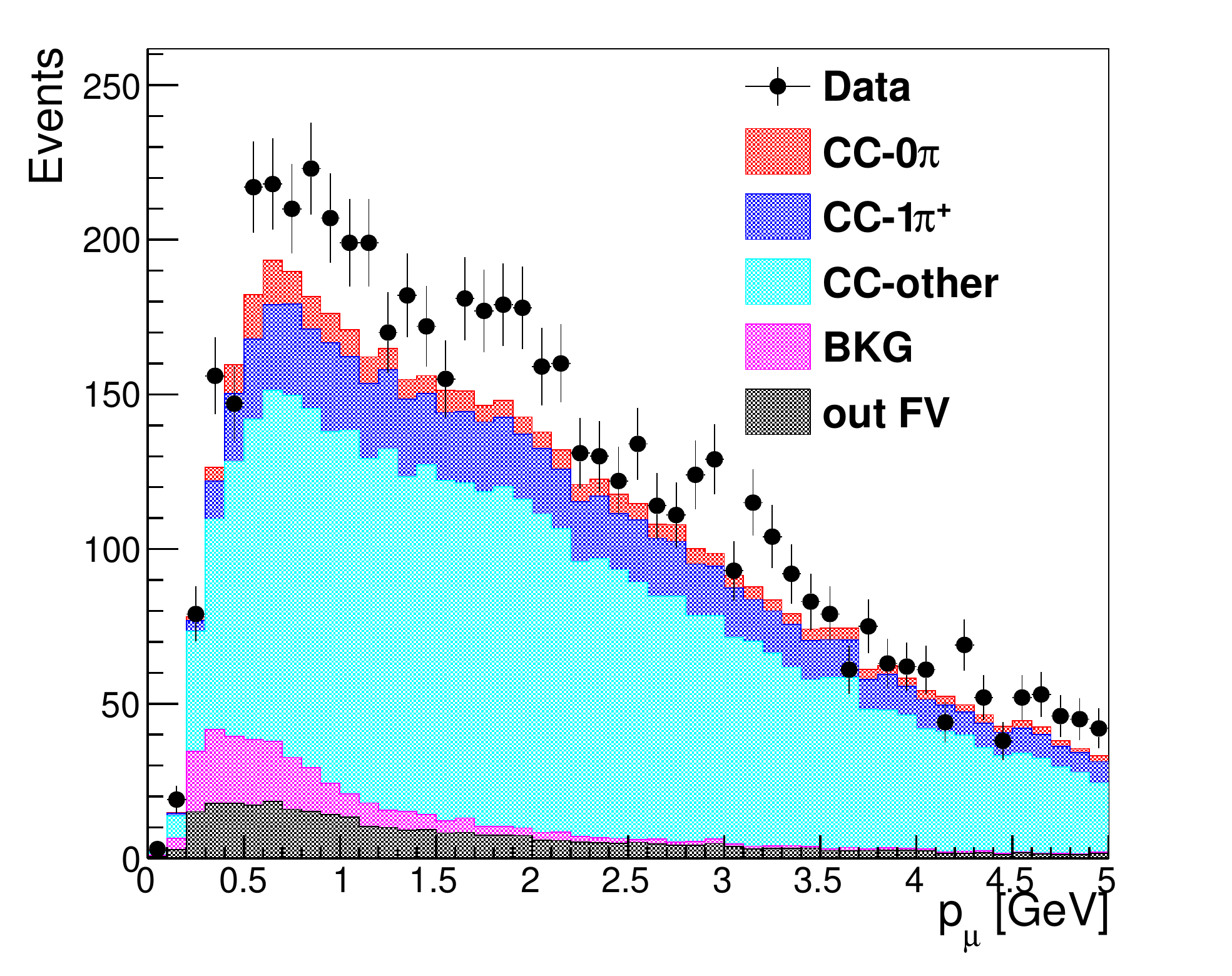}
  \caption[Data and MC distributions of the CCOther selection]{Data
    and MC distributions of the water-in CCOther sideband selection. NEUT
    5.3.2 MC has been normalized to data POT and is sorted into
    various truth topologies.}
\label{fig:ccothersel}
\end{figure}

To provide a data-driven constraint on the background, %additional    % meant as ``additional selections'' but confusing as ``additional sidebands'', so best removed to avoid confusion.
sideband selections are used for the \ccopi{} and CCOther
topologies. Their selection is identical to the signal selection with
the exception of the final cut. For the \ccopi{} sideband, events with
exactly two \pd{} reconstructed tracks in a bunch along with at least
one \pd{} reconstructed Michel electron are selected. For the CCOther
sideband, events with greater than two \pd{} reconstructed tracks are
selected. These cuts reduce overlap between the signal and sideband
selections as shown in \cref{fig:cc1pisel,fig:ccothersel} for the water-in samples.  The water-out sideband samples look very similar.  The numbers of selected events in the signal and sidebands regions are summarized in Tables~\ref{tab:cutsair} and~\ref{tab:cutsh2o}.

\begin{table}
  \begin{ruledtabular}
  \begin{tabular}{lrrr}
            &  \multicolumn{3}{c}{Water-Out Data} \\	
           &  \multicolumn{3}{c}{Num of Selected Events} \\	
    Cut & \cczpi{}  & \ccopi{} & CC Oth \\
    \hline
    Data Quality and \pd{}+TPC  \\
    \hspace{0.1in} $\mu^{-}$ candidate in Fid. Vol. & 33083 & 33083 & 33083 \\
    \# \pd{} Tracks & 13370 & 8107 & 11606 \\
    \# Michel e$^{-}$ & NA & 1710 &  NA \\
  \end{tabular}
\end{ruledtabular}
  \caption[Cuts on Water-Out Data]{Number of water-out data events contained in the selected signal and sideband samples after each selection cut is applied.  The cut on the number of Michel electrons is only applied to the \ccopi{} sample.}
  \label{tab:cutsair}
\end{table}

\begin{table}
  \begin{ruledtabular}
\begin{tabular}{lrrr}
            &  \multicolumn{3}{c}{Water-In Data} \\	
           &  \multicolumn{3}{c}{Num of Selected Events} \\	
    Cut & \cczpi{}  &  \ccopi{} & CC Oth \\
    \hline
    Data Quality and \pd{}+TPC  \\
    \hspace{0.1in} $\mu^{-}$ candidate in Fid. Vol. & 27713 & 27713 & 27713 \\
    \# \pd{} Tracks & 12777 & 6297 & 8639 \\
    \# Michel e$^{-}$ & NA & 1434 &  NA \\
  \end{tabular}
\end{ruledtabular}
  \caption[Cuts on Water-In Data]{Number of water-in data events contained in the selected signal and sideband samples after each selection cut is applied. The cut on the number of Michel electrons is only applied to the \ccopi{} sample.}
  \label{tab:cutsh2o}
\end{table}

\subsection{Cross Section Extraction}
The event selections described above are binned in the reconstructed
double differential $(p_\mu, \cos \theta_\mu)$ phase space.
Reconstructed kinematics are approximations to the true initial state
of the muon.   % an approximation is always imperfect        // avoid 's 
To extract the true kinematics from the
reconstructed, an unfolding technique is used based on D'Agostini's
method with the MC truth as the
prior~\cite{dagostini1995bayesunf}. The purpose of unfolding is to        %% This sentence should be rewritten without ``remove''
remove detector reconstruction related imperfections to achieve a more
accurate representation of how the muon emerged from the
interaction. By correcting the data samples from \pd{} water-in and water-out configurations
separately, a subtraction procedure ultimately gives a cross section
on water.  Additionally, the MC is tuned to account for flux, interaction
modeling, and detector corrections. The final tuned MC is then used to
calculate efficiency corrections, purity corrections, and the
unfolding matrix.

The exclusive \cczpi{} signal is evaluated by correcting the data
selection based on the \cczpi{} signal purity in the MC\@. The signal purity is calculated using the MC truth information and the data from the
sideband samples. As described above, \ccopi{} interactions are the
largest sources of background, followed by CCOther
interactions. Therefore, sideband samples for these two backgrounds
were selected as a data-driven background constraint. The                %% Additional constraint? Removed..
ratio of the overall data sideband normalization to the overall MC
sideband normalization is calculated and used to constrain the
corresponding background in the \cczpi{} MC selection as, 
\begin{equation}
  \label{eq:bprime}
  B_j' = B_j \frac{S_d}{S_m},
\end{equation}
where $j$ denotes a bin index, $B'$ the sideband constrained
background, $B$ the original background in the MC signal selection,
and $S_d$ and $S_m$ the sideband normalizations from data and MC
respectively. This affects the signal purity used in the background
correction,
\begin{equation}
  \label{eq:pprime}
  p_j' = 1 - \frac{B_j'}{N_{m,j}'},
\end{equation}
where $N_{m,j}' = N_{m,j}-B_j+B_j'$ and $N_m$ is the original number
of events in the \cczpi{} MC selection.

%% Suggest to remove the whole paragraph: PRD readers know what unfolding is.
% It is easiest to think about the unfolding in terms of two main
% categories, true kinematics (causes) and reconstructed kinematics
% (effects). Imperfect reconstruction will smear the correspondence between
% cause and effect bins. In order to recapture the true distribution, a
% probabilistic approach is taken. First, $\Pr(E_j \mid C_i)$, the
% probability of an event in cause bin $i$ migrating to effect bin $j$,
% is calculated from the MC\@. By Bayes' theorem the posterior can be
% calculated as
% \begin{equation}
%   \label{eq:bayes}
%   \Pr(C_i \mid E_j) \propto \Pr(E_j \mid C_i) \Pr(C_i)
% \end{equation}
% to give the conditional probability of an event falling into true bin
% $i$ given that it was measured in effect bin $j$. Once the posterior
% probability is calculated for each bin it can be used to unfold the
% reconstructed distribution. The strategy outlined by D'Agostini is
% then to iterate this step and use the newly calculated true
% distribution as a new prior $\Pr(C_i)$. 
For this analysis, the initial prior used in the D'Agostini unfolding
technique was taken from the MC truth and a single iteration is
used. Fake-data studies showed that multiple iterations did not
improve results but increased uncertainties.  This increase is due to
the correlation introduced between the data and successive
priors~\cite{adye_unfolding} and can produce large fluctuations between neighboring bins.  Additionally while in fake data tests the total $\chi^{2}$ for the unfolded result was slightly larger than expected ($\sim$1.7 not including any systematics), it did not significantly decrease with additional iterations.  Thus a single iteration was chosen.  This must be regarded as a regularized result.   With the MC truth as the prior, a single
iteration Bayesian unfolding matrix is equivalent to directly
constructing an unfolding matrix based on the MC\@. This means that
the unfolding matrix is calculated based on the mapping between truth
and reconstructed kinematics in the MC\@.  

%To extract a measurement on water requires an additional subtraction
To extract the neutrino cross section on water an additional
subtraction step after the unfolding is required. The \pd{} fiducial
volume contains plastic scintillator layers and thin brass sheets
sandwiched between layers of water~\cite{p0dnim}. The water layers act
as a passive target making it difficult to know whether an interaction
occurred on water or on some other target nucleus. To work around
this, the \pd{} was designed to be drained and filled during different
run periods. Everything being equal except for the inclusion or
exclusion of water, a subtraction of the true water-in and water-out
distributions should give the number of interactions on water. Since
the \pd{} is in two different detector configurations with and without
water, it is necessary to correct for detector smearing and
inefficiencies before the subtraction. A direct subtraction of the
reconstructed event rates would give an incorrect estimate for the
actual event rate on water. Therefore, we first unfold the
reconstructed distribution for water-in and water-out separately to
get an approximation of their true distributions, then subtract the
unfolded distributions to get the distribution of interactions on
water. Specifically, this is given by,
\begin{equation}
  \label{eq:unfoldsubtract}
  N_{i}^{\ce{H2O}} = \frac{U_{ij}^{w} N_{j}^{w}}{\epsilon_{i}^{w}} - R
  \frac{U_{ij}^{a} N_{j}^{a}}{\epsilon_{i}^{a}},
\end{equation}
where the indexes $i$ and $j$ indicate true and reconstructed bins
respectively, $w$ and $a$ indicate water-in and water-out periods
respectively, $N$ is the number of purity-corrected, signal events
measured in the data signal selection, $\epsilon$ the selection
efficiency, and $R$ the flux normalization factor between water-in and
water-out periods. $U_{ij}$ represents the unfolding matrix. From
this, the differential cross section on water can be expressed as,
\begin{equation}
  \label{eq:diffxsec}
  \frac{d\sigma}{dx_i} = \frac{N_{i}^{\ce{H2O}}}{F^{w} N_{n} \Delta x_i},
\end{equation}
where $F^{w}$ is the integrated flux over the water-in period, $N_{n}$
the number of neutrons, and $\Delta x_i$ the area of bin $i$ across
the phase space $x$. $F^w$ is the POT-weighted flux of all water-in
periods and calculated using flux simulations and from the constraints
described above. As the water-in and water-out periods have different
beam exposures, total flux for the water-out periods, $F^a$, is used
to scale the flux normalization ratio, $R = F^w/F^a$. Additionally,
correlated errors between the water-in and water-out periods are taken
into account in the error propagation and discussed in
\cref{sec:systs}.

\subsection{Systematic Uncertainties}
\label{sec:systs}

As in previous T2K analyses, the systematic uncertainties can be separated into three principal categories: flux uncertainties, cross section model uncertainties, and detector uncertainties~\cite{t2k_ccqe,t2k_oscillation}.

The flux and cross section errors are calculated by reweighting the individual simulated events. 
Once the events are reweighted, the unfolding matrix is regenerated and the cross section is recalculated using the new weights.
The uncertainties for flux and cross section parameters are taken from the same covariance matrices used as inputs to the near detector fit in the T2K oscillation analysis~\cite{t2k_oscillation}.   There are 46 total parameters in the covariance matrix: 25 for flux, 6 for final-state interactions, and 15 other cross section parameters.
By generating many throws for the model parameters and calculating the resulting event weights, we can calculate a covariance matrix for the final result. If $\frac{d\sigma_O}{dx_{i;k}}$ gives the cross section result for bin $i$ and parameter throw $k$ which has $n$ throws, then the covariance between bins $i$ and $j$ is
\begin{equation}
\label{eq:cov}
\mathrm{cov}(i,j) = \frac{1}{n}\sum\limits_{k=0}^{n-1} \left(\frac{d\sigma_O}{dx_{i;k}} - \bar{\frac{d\sigma_O}{dx_i}}\right)\left(\frac{d\sigma_O}{dx_{j;k}} - \bar{\frac{d\sigma_O}{dx_j}}\right)
\end{equation}
where
\begin{equation}
\bar{\frac{d\sigma_O}{dx_i}} = \frac{1}{n}\sum\limits_{k=0}^{n-1} \frac{d\sigma_O}{dx_{i;k}}
\end{equation}

The flux uncertainty is dominated by the uncertainty in the total flux
normalization and is found to be 8.76\%. Flux uncertainties are due in
large part to uncertainties in the hadron production model but are
affected by beamline uncertainties as well~\cite{t2k_flux}. A
parameterization in neutrino energy and flavor is used to propagate
flux uncertainties. Cross section model uncertainties include both
uncertainties in basic neutrino-nucleus scattering and uncertainties
from FSI~\cite{t2k_oscillation}. Parameters that govern the neutrino
interaction and nuclear description models used in the NEUT
generator~\cite{neut} are used to propagate cross section
uncertainties.

%TODO: talk about NEUT, GENIE, $M_A = \SI{1.15 \pm 0.03}{\GeV / c^2}$,
% $\text{2p2h norm.} = \SI{27 \pm 12}{\percent}$, and $p_F = \SI{223 \pm
%   5}{\MeV / c}$
Neutrino-nucleus scattering depends on nucleon form factors and the
nuclear medium model. Neutrino generators typically assume a dipole
form factor governed by an axial mass parameter for QE and resonant
charged-current interactions. Additionally, many of the interaction
parameters tune the normalization on certain interaction channels that
are poorly understood. These include 2p2h contributions
that arise from multinucleon correlations inside the nucleus. The
current understanding of these multinucleon interactions is based on
the meson exchange model where two nucleons exchange a meson current
such that the charged-current interaction involves both nucleons. This
is expected to enhance the CCQE-like cross section. Thus, instead of a
single final state nucleon, two or more nucleons may be ejected out of
the nucleus. Other parameters, such as the nuclear binding energy and
the Fermi momentum describe the nuclear medium. 
%A covariance matrix that describes the correlations between these parameters is used to throw and reweight the MC as shown in \cref{eq:cov}.

An intranuclear cascade model describes secondary pion propagation
within the nucleus~\cite{fsi}. Uncertainties on FSI are calculated by
tuning the pion production, absorption, rescattering and
charge-exchange probabilities.

The primary neutrino generator used for this analysis is a tuned
version of NEUT 5.3.2~\cite{neut}~\cite{hayato}.The tuning is applied based on
best-fit values from a fit of three models implemented in NEUT to
external \minerva{} and \miniboone{} CCQE
measurements~\cite{wilkinson}. The NEUT model uses the Smith-Moniz
relativistic Fermi gas (RFG) model~\cite{rfg} with a relativistic random phase approximation (RPA)~\cite{rpa} and a multinucleon exchange (2p2h) model~\cite{rpa}~\cite{2p2h}. An axial mass of $M_A^{QE} = \SI{1.15 \pm 0.03}{\GeV / c^2}$, the amplitude of the multinucleon effects 
$\text{2p2h norm.} = \SI{27 \pm 12}{\percent}$, and  Fermi momentum $p_F = \SI{223 \pm
  5}{\MeV / c}$ were used. The uncertainties in these parameters are propagated
to the final cross section measurement. A secondary neutrino generator
used for testing is GENIE 2.8.0~\cite{genie}.

\Cref{tab:overall_xsec_errors} shows the individual contributions of
several interaction model parameters to the total cross section. Note
that the 2p2h normalization, binding energy $E_b$, and Fermi momentum
$p_F$ are nuclei-dependent. These are calculated with \num{100} throws
without taking correlations into account. This illustrates
and identifies the effects of a single parameter on the cross section
result. For the final result, 800 throws were used along
with covariances across the full parameter-space to generate the
interaction and other systematic uncertainties, and 2000 throws were used for statistical errors. The lower number of throws used in Table~\ref{tab:overall_xsec_errors} means
there is a larger statistical error on the uncertainty itself.

\begin{table}[htb]
  \begin{ruledtabular}
  \begin{tabular}{lr}
    Source & Uncertainty [\%]\\
    &  on cross sec. \\
    \hline
    $M_A^{QE}$ & \num{0.8}\\ 
    $M_A^{RES}$ & \num{0.3}\\ 
    CA5 single-pion interaction & \num{0.5}\\ 
    2p2h normalization (\ce{^{16}O}) & \num{2.8}\\ 
    $E_b$ (\ce{^{16}O}) & \num{0.05}\\ 
    $p_F$ (\ce{^{16}O}) & \num{0.2}\\ 
    % $M_A^{QE}$ & \num{0.84}\\ 
    % $M_A^{RES}$ & \num{0.29}\\ 
    % CA5 single-pion interaction & \num{0.46}\\ 
    % 2p2h normalization (\ce{^{16}O}) & \num{2.83}\\ 
    % $E_b$ (\ce{^{16}O}) & \num{0.05}\\ 
    % $p_F$ (\ce{^{16}O}) & \num{0.20}\\ 
  \end{tabular}
  \end{ruledtabular}

  \caption[Fractional uncertainties from individual cross section
  parameters]{The fractional uncertainties on the restricted phase space cross
    section due to several of the individual cross section parameters.}
  \label{tab:overall_xsec_errors}
\end{table}

Detector systematics describe uncertainties in reconstructed properties of events rather than the underlying physics. These are separated into weight systematics and variation systematics. Weight systematics modify the weight to be applied to simulated events. These include effects such as differences in tracking efficiencies between data and simulation. Variation systematics modify observables such as the reconstructed particle momentum of a simulated event. These allow for migration of events between different kinematic bins. Systematics related to the uncertainty on the fiducial mass must be treated in a different manner, as these affect both the expected number of events and the total mass used to normalize the cross section measurement.

The event reconstruction in the \pd{} uses a Kalman filter algorithm to calculate the properties of tracks found in the detector. 
The reconstructed momentum will primarily be influenced by two main effects: the curvature of the track in the tracker and the energy loss within the \pd{}. 
Because relatively little information about the track curvature can be obtained by the \pd{} and because the small amount of energy loss in the tracker will have little effect on the reconstructed initial momentum, the systematics for these two effects are treated separately. 
Uncertainties in the energy loss in the \pd{} are estimated from a sample of cosmic ray muons that pass through the tracker and stop in the \pd{}. 
To propagate these uncertainties, the reconstructed momentum in the simulation is altered by
\begin{equation}
(\Delta p)^{\pd{}}_{scale} = (x_s \sigma_s) (p^{reco} - p^{reco}_{TPC1})
\end{equation}
for the scale uncertainty and
\begin{equation}
(\Delta p)^{\pd{}}_{res} = (x_r \sigma_r) (p^{reco} - p^{reco}_{TPC1} - p^{truth} + p^{truth}_{TPC1})
\end{equation}
for the resolution uncertainty. In these equations $\sigma$ represents the uncertainty and $x$ is a normally distributed random number that changes for each systematics throw.
Similar numbers are obtained for both the water-out and water-in geometries, and a 1.4\% uncertainty is applied to the energy loss scale and a 7\% uncertainty to the resolution. 
Tracker curvature systematics follow the same procedure as for previous T2K analyses~\cite{t2k_oscillation}. 
Scale and resolution uncertainties are applied to the momentum extrapolated from the track curvature. 
Additional uncertainties are applied to account for distortions in the magnetic field, charge misidentification, TPC cluster reconstruction, and TPC track reconstruction. 
Studies have shown that the \pd{} track reconstruction and matching \pd{} tracks to TPC tracks have an efficiency greater than 99.8\%, so no correction or uncertainty is applied due to this.

In some cases, an interaction that occurred outside the fiducial volume will be reconstructed in the fiducial volume. Using the nominal NEUT simulation, such out of fiducial volume events represent 2.89\% of events in the water-in sample and 3.95\% of events in the water-out sample. An analysis of the spatial distribution of events in the layers of the \pd{} central electromagnetic calorimeter, immediately downstream of the water target, suggests that there may be as much as a factor of two discrepancy in the amount of migration between data and simulation (but this is a small fraction of the total number of events). To account for this, we increase the weights of events originating in the central electromagnetic calorimeter by a factor of two and add an uncertainty of 50\% of the initial weight. The distributions of these events in the two \pd{} geometries are identical, as there is no water in these layers. These events will cancel out up to small differences in statistics and the detector response, so a large uncertainty corresponds to a small effect on the final result.  The detector systematics are summarized in Table~\ref{table:DetSystematics}

\begin{table}
\begin{ruledtabular}
\begin{tabular}{lr}
Systematic & Uncertainty [\%]\\
\hline
\pd{} Energy Loss Scale  & 1.3 \\
\pd{} Energy Loss Resolution & 6.7\\
Tracker Momentum Scale & 1.5\\
Tracker Momentum Resolution & 0.2 \\
Magnetic Field Distortion  & 0.04\\
Charge ID Efficiency & 0.1\\
TPC Cluster Efficiency  & 0.3\\
TPC Track Efficiency & 0.4\\
Out of Fiducial Volume &  0.8\\
Detector Mass &  1.5 \\
% \pd{} Energy Loss Scale & Variation  & 1.34 \\
% \pd{} Energy Loss Resolution & Variation & 6.72\\
% Tracker Momentum Scale & Variation & 1.49\\
% Tracker Momentum Resolution & Variation & 0.19 \\
% Magnetic Field Distortion & Variation & 0.04\\
% Charge ID Efficiency & Weight & 0.13\\
% TPC Cluster Efficiency & Weight & 0.29\\
% TPC Track Efficiency & Weight & 0.44\\
% Out of Fiducial Volume & Weight & 0.76\\
% Detector Mass & Mass & 1.46 \\
\end{tabular}
\end{ruledtabular}
\caption{Detector systematic uncertainties considered in this analysis. The uncertainty in each category gives the overall uncertainty on the restricted phase space cross section, calculated for 100 throws. The final result in~\cref{eq:reducedxsec} throws all uncertainties simultaneously and uses a larger number of throws.}
\label{table:DetSystematics}
\end{table}

The one-track selection is sensitive to pileup where particles from several interactions are reconstructed in the \pd{} in the same bunch.
The main pileup sources are cosmic ray muons, multiple neutrino interactions in the same beam bunch, and interactions occurring in the material outside ND280 (called sand events) that produce tracks in the \pd{}. The cosmic ray muon pileup rate is estimated to be approximately $2\times10^{-4}$ per bunch, which is negligible. The beam-related pileup rate is dominated by sand events, and is calculated separately for each ND280 run period. The sand muon corrections are given in \cref{table:Pileup}.

\begin{table}
\begin{ruledtabular}
\begin{tabular}{lr}
Run & $p_{sand}$ [bunch$^{-1}$] \\
\hline
% 1 & 0.0085 \\
2 water & 0.0144\\
2 air & 0.0173 \\
%3b & 0.0158 \\
3c & 0.0188 \\
4 water & 0.0217 \\
4 air & 0.0245 \\
\end{tabular}
\end{ruledtabular}
\caption{Sand event pileup rates for each ND280 run period. The weights of simulated events are reduced by a factor of $1-p_{sand}$. }
\label{table:Pileup}
\end{table}

Finally, there is some uncertainty in the total target mass. Mass
uncertainties change the total number of events expected as well as
the total number of target nuclei used to obtain the correct cross
section normalization. Because the \pd{} is composed of several
materials, the mass uncertainty also has a small effect on the
smearing matrix, since the cross section shape will not necessarily
have the exact same shape for different materials. In the case of the
water in the fiducial volume, the total mass of the water is known
very well, but there is still uncertainty in how the water is
distributed throughout the different layers. Measurements have shown
that there is approximately 15\% more water in the most downstream
water layer compared to more centrally located layers. There is also a
smaller uncertainty on the amount of water in individual layers away
from the edges.  To estimate the uncertainty due to the mass, a 0.8\%
uncertainty on the total water mass is applied with additional 15\%
uncertainties in the layers at the ends of the water target and 5\%
uncertainties on the remaining layers. The layer-by-layer variations
are performed keeping the total mass fixed. These give us an
uncertainty of 1.5\% due to the target mass.

For all systematic errors, correlations between the water-in and
water-out periods were taken into consderation in
\cref{eq:unfoldsubtract}. This was done by ensuring identical seeding
for throws in calculating \cref{eq:cov}. Since flux uncertainties do
not distinguish between water-in and water-out samples the flux
uncertainties should be fully correlated. The oxygen binding energy
parameter affects only the water-in sample, which means that the
overall cross section uncertainties are not fully correlated. Some
\pd{}-specific detector systematics, such as the \pd{} target mass
systematic, have dependence on the water state. Thus detector
systematics are not fully correlated either. Further, any differences
between water-in and water-out detectors are taken into account with
different standard deviations on the underlying
variations. Statistical errors are treated as uncorrelated.

\section{\label{sec:results}Results}
\begin{figure*}
  \centering
  \subfloat{
    \includegraphics[width=0.5\textwidth,trim={0 7cm 13.3cm 0cm},clip]{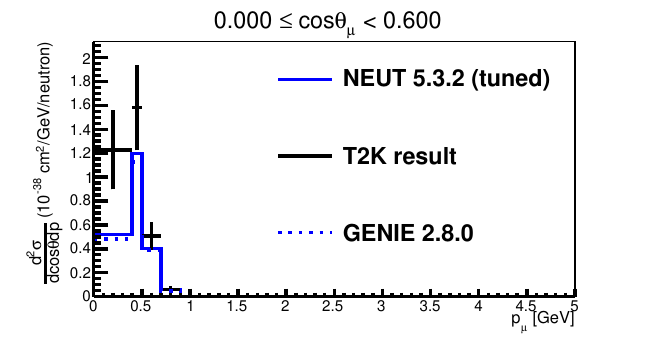}
  }
  \subfloat{
    \includegraphics[width=0.5\textwidth,trim={6.65cm 7cm 6.65cm 0cm},clip]{resultmasterxsec}
  }\\
  \subfloat{
    \includegraphics[width=0.5\textwidth,trim={13.3cm 7cm 0cm 0cm},clip]{resultmasterxsec}
  }
  \subfloat{
    \includegraphics[width=0.5\textwidth,trim={0cm 3.5cm 13.3cm 3.5cm},clip]{resultmasterxsec}
  }\\
  \subfloat{
    \includegraphics[width=0.5\textwidth,trim={6.65cm 3.5cm 6.65cm 3.5cm},clip]{resultmasterxsec}
  }
  \subfloat{
    \includegraphics[width=0.5\textwidth,trim={13.3cm 3.5cm 0cm 3.5cm},clip]{resultmasterxsec}
  }\\
  \subfloat{
    \includegraphics[width=0.5\textwidth,trim={0cm 0cm 13.3cm 7cm},clip]{resultmasterxsec}
  }
  \subfloat{
    \includegraphics[width=0.5\textwidth,trim={6.65cm 0cm 6.65cm 7cm},clip]{resultmasterxsec}
  }\\
  \caption[The double differential \numu{} \cczpi{} cross section]{The
    double differential \numu{} \cczpi{} cross section on water for
    each slice in $\cos \theta_\mu$.}
   \label{fig:result_master_xsec}
\end{figure*}

The result shown in \cref{fig:result_master_xsec} uses data from T2K
Runs \numrange{2}{4}. It is reported as a double differential cross
section in the outgoing muon kinematics $(p_\mu,\cos\theta_\mu)$. Black data points taken from~\cref{eq:diffxsec} show the double differential result with
full error bars. The
colored error bars show the cumulative uncertainties from various
sources, starting with the data statistics and ending with the
detector systematics. Errors from each source are added in
quadrature. MC predictions from NEUT 5.3.2 (tuned) and GENIE 2.8.0
are shown as solid and dashed blue lines.
%INCLUDE? Tables containing the full result and errors
% can be found in \cref{sec:details}.

\begin{figure*}
  \centering
  \subfloat{
    \includegraphics[width=0.5\textwidth,trim={0 7cm 13.3cm 0cm},clip]{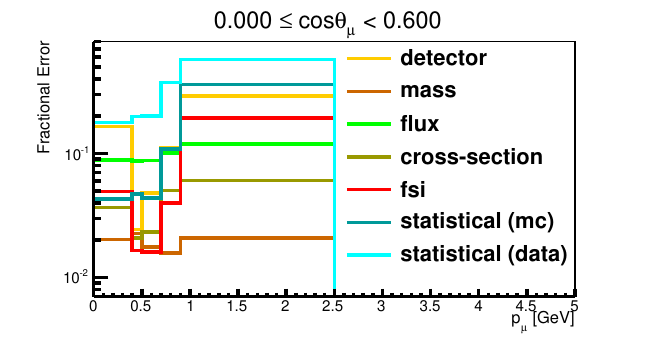}
  }
  \subfloat{
    \includegraphics[width=0.5\textwidth,trim={6.65cm 7cm 6.65cm 0cm},clip]{resultmastererr}
  }\\
  \subfloat{
    \includegraphics[width=0.5\textwidth,trim={13.3cm 7cm 0cm 0cm},clip]{resultmastererr}
  }
  \subfloat{
    \includegraphics[width=0.5\textwidth,trim={0cm 3.5cm 13.3cm 3.5cm},clip]{resultmastererr}
  }\\
  \subfloat{
    \includegraphics[width=0.5\textwidth,trim={6.65cm 3.5cm 6.65cm 3.5cm},clip]{resultmastererr}
  }
  \subfloat{
    \includegraphics[width=0.5\textwidth,trim={13.3cm 3.5cm 0cm 3.5cm},clip]{resultmastererr}
  }\\
  \subfloat{
    \includegraphics[width=0.5\textwidth,trim={0cm 0cm 13.3cm 7cm},clip]{resultmastererr}
  }
  \subfloat{
    \includegraphics[width=0.5\textwidth,trim={6.65cm 0cm 6.65cm 7cm},clip]{resultmastererr}
  }\\
  \caption[The fractional error on the double differential result]{The
    fractional error from each source of uncertainty on the double
    differential \numu{} \cczpi{} cross section on water.}
  \label{fig:result_master_err}
\end{figure*}

The fractional contribution from each source of uncertainty is shown
in \cref{fig:result_master_err}, binned and plotted in the same scheme
as \cref{fig:result_master_xsec}. In most regions, the statistical
error from the data is the single most dominant source of bin-by-bin
uncertainty, but overall the statistical errors are comparable to the total systematic uncertainty. Aside from the low sensitivity bins, the
fractional bin-by-bin errors lie on the order of
\SIrange{10}{20}{\percent}.

\begin{figure*}
  \centering
  \subfloat{
    \includegraphics[width=0.5\textwidth,trim={0 7cm 13.3cm 0cm},clip]{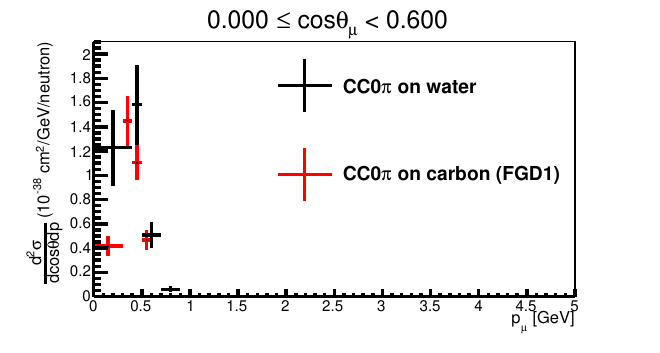}
  }
  \subfloat{
    \includegraphics[width=0.5\textwidth,trim={6.65cm 7cm 6.65cm 0cm},clip]{compareresultfgd1}
  }\\
  \subfloat{
    \includegraphics[width=0.5\textwidth,trim={13.3cm 7cm 0cm 0cm},clip]{compareresultfgd1}
  }
  \subfloat{
    \includegraphics[width=0.5\textwidth,trim={0cm 3.5cm 13.3cm 3.5cm},clip]{compareresultfgd1}
  }\\
  \subfloat{
    \includegraphics[width=0.5\textwidth,trim={6.65cm 3.5cm 6.65cm 3.5cm},clip]{compareresultfgd1}
  }
  \subfloat{
    \includegraphics[width=0.5\textwidth,trim={13.3cm 3.5cm 0cm 3.5cm},clip]{compareresultfgd1}
  }\\
  \subfloat{
    \includegraphics[width=0.5\textwidth,trim={0cm 0cm 13.3cm 7cm},clip]{compareresultfgd1}
  }
  \subfloat{
    \includegraphics[width=0.5\textwidth,trim={6.65cm 0cm 6.65cm 7cm},clip]{compareresultfgd1}
  }\\
  \caption[Comparison against the T2K \cczpi{} cross section on
  \ce{C8H8}]{Here, the \cczpi{} water cross section from the \pd{} is
    overlaid with the \cczpi{} \ce{C8H8} cross section previously
    measured by T2K~\cite{t2k_ccqe}. The error bars shown include all
    sources of uncertainty \emph{except} the flux. In regions where
    $\cos \theta_\mu$ slices are different between the water and
    hydrocarbon result, the bin edges are noted in the legend.}
  \label{fig:compare_result_fgd1}
\end{figure*}

This result is compared to the T2K \cczpi{} cross section on
\ce{C8H8}~\cite{t2k_ccqe}  in
\cref{fig:compare_result_fgd1}. Since both results were obtained using
the T2K flux, the error bars shown do not include the flux
uncertainty. The binnings do not exactly match between the two
analyses, but when overlaid, both results appear to be consistent
within uncertainties for most of the phase space. The largest areas of
discrepancy are in the high-angle regions, where the \cczpi{} water
cross section is higher than the hydrocarbon cross section.

\begin{figure*}
  \centering
  \subfloat{
    \includegraphics[width=0.5\textwidth,trim={0 7cm 13.3cm 0cm},clip]{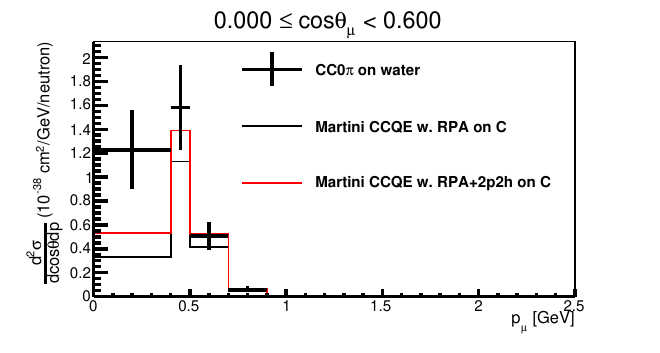}
  }
  \subfloat{
    \includegraphics[width=0.5\textwidth,trim={6.65cm 7cm 6.65cm 0cm},clip]{compareresultmartini}
  }\\
  \subfloat{
    \includegraphics[width=0.5\textwidth,trim={13.3cm 7cm 0cm 0cm},clip]{compareresultmartini}
  }
  \subfloat{
    \includegraphics[width=0.5\textwidth,trim={0cm 3.5cm 13.3cm 3.5cm},clip]{compareresultmartini}
  }\\
  \subfloat{
    \includegraphics[width=0.5\textwidth,trim={6.65cm 3.5cm 6.65cm 3.5cm},clip]{compareresultmartini}
  }
  \subfloat{
    \includegraphics[width=0.5\textwidth,trim={13.3cm 3.5cm 0cm 3.5cm},clip]{compareresultmartini}
  }\\
  \subfloat{
    \includegraphics[width=0.5\textwidth,trim={0cm 0cm 13.3cm 7cm},clip]{compareresultmartini}
  }
  \subfloat{
    \includegraphics[width=0.5\textwidth,trim={6.65cm 0cm 6.65cm 7cm},clip]{compareresultmartini}
  }\\
  \caption[Comparison against the Martini model]{A comparison of the
    \cczpi{} water cross section against two Martini model predictions
    on carbon, one with 2p2h contributions and one without.}
  \label{fig:compare_result_martini}
\end{figure*}

\begin{figure*}
  \centering
  \subfloat{
    \includegraphics[width=0.47\textwidth]{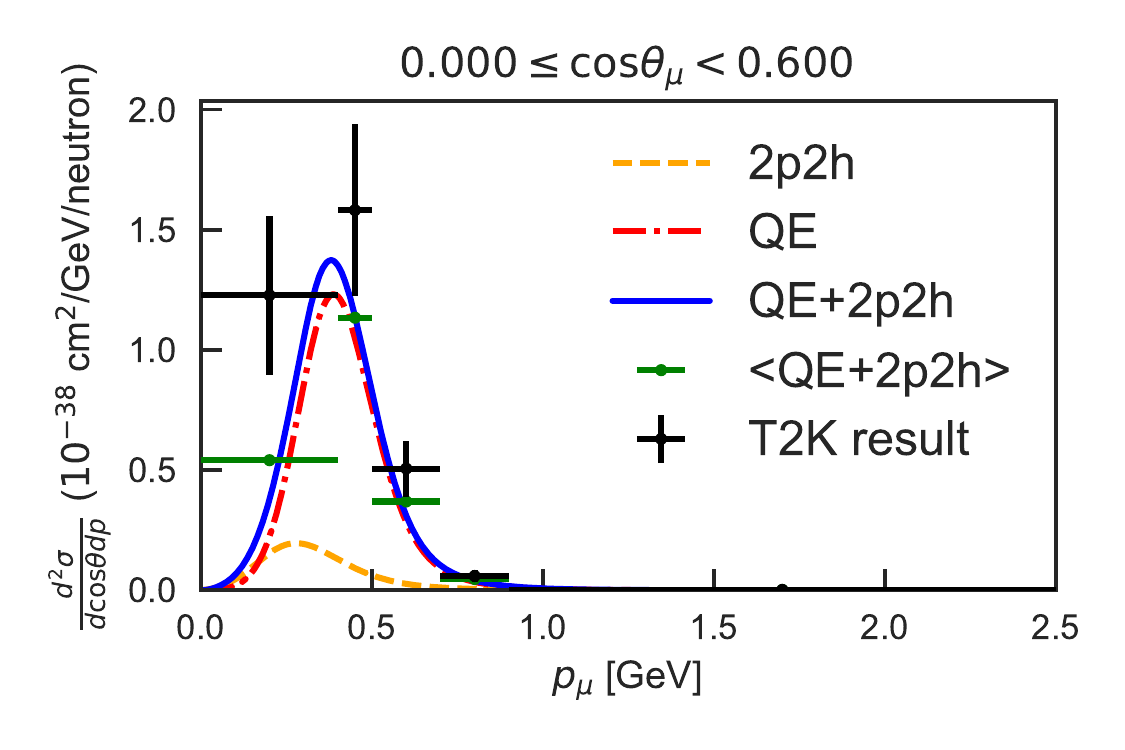}
  }
  \subfloat{
    \includegraphics[width=0.47\textwidth]{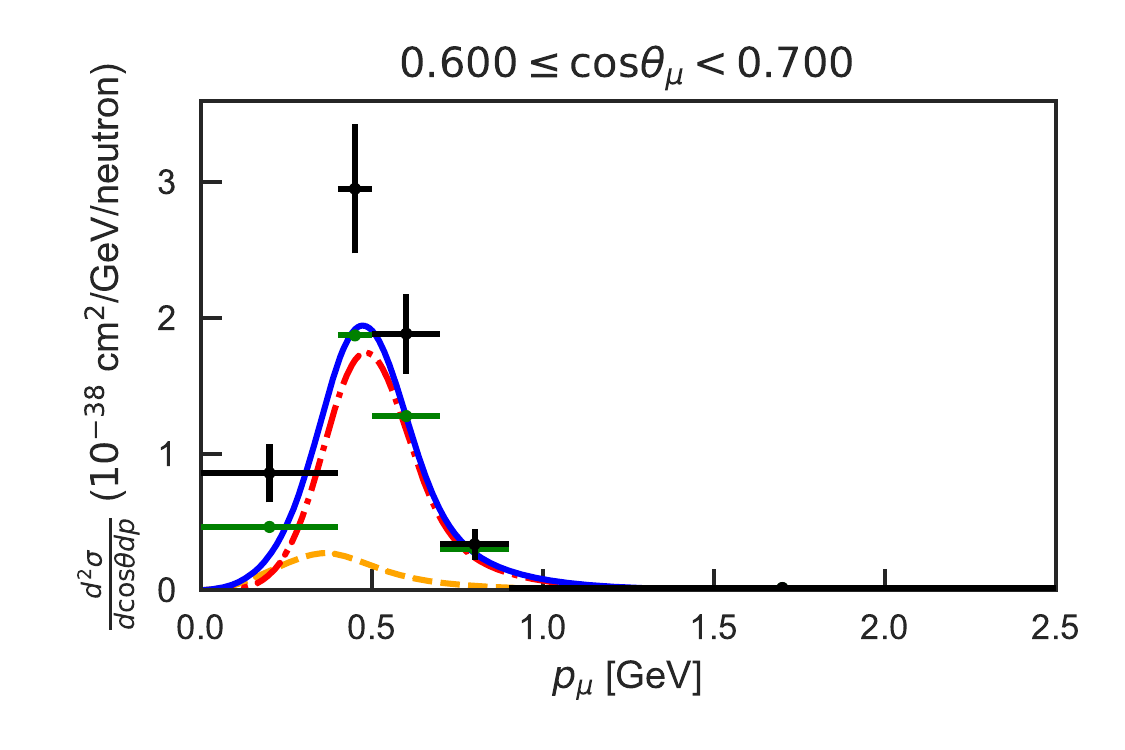}
  }\\
  \subfloat{
    \includegraphics[width=0.47\textwidth]{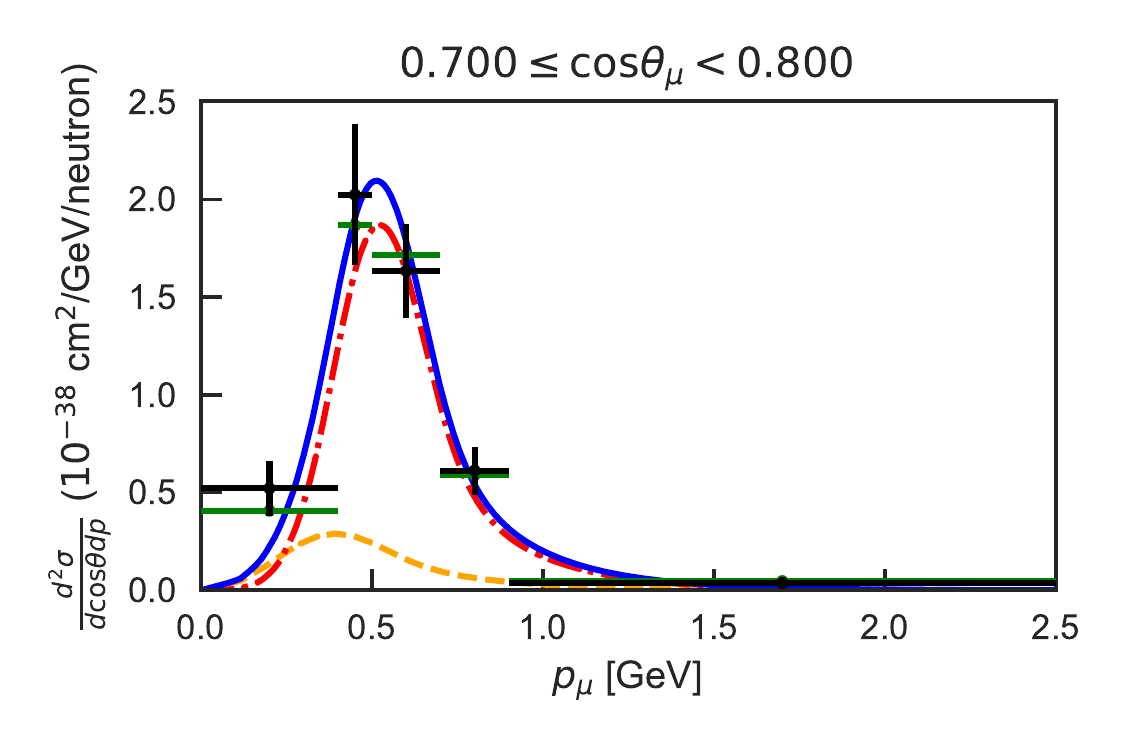}
  }
  \subfloat{
    \includegraphics[width=0.47\textwidth]{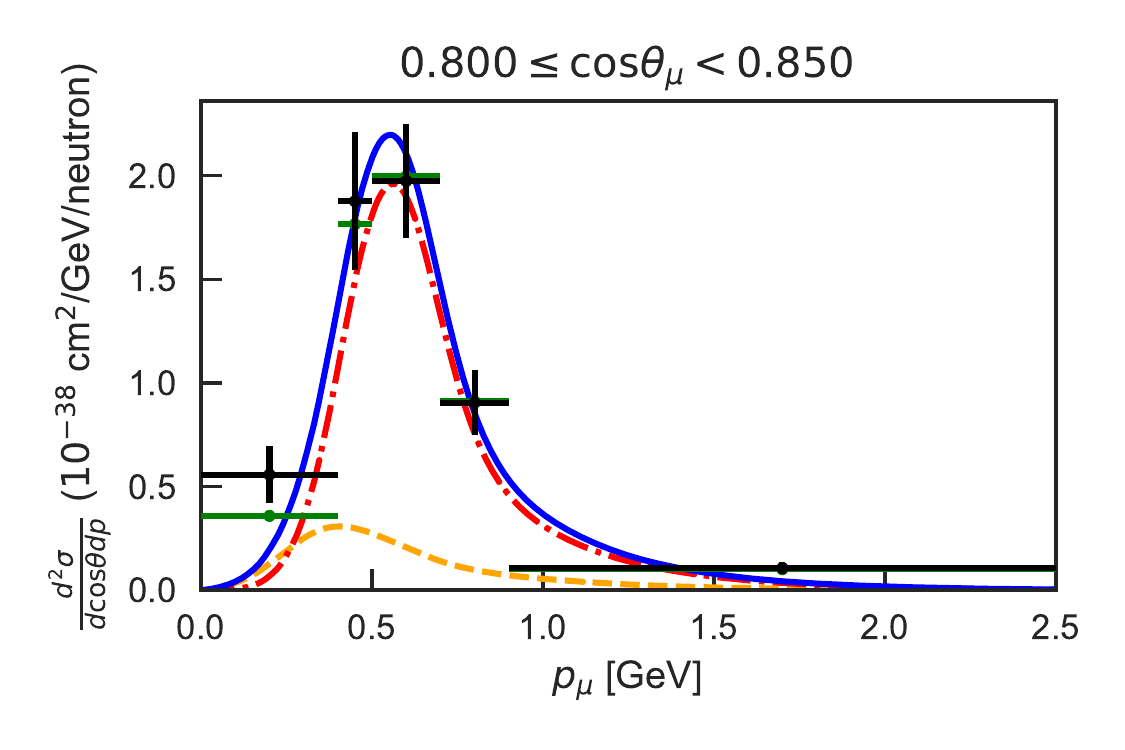}
  }\\
  \subfloat{
    \includegraphics[width=0.47\textwidth]{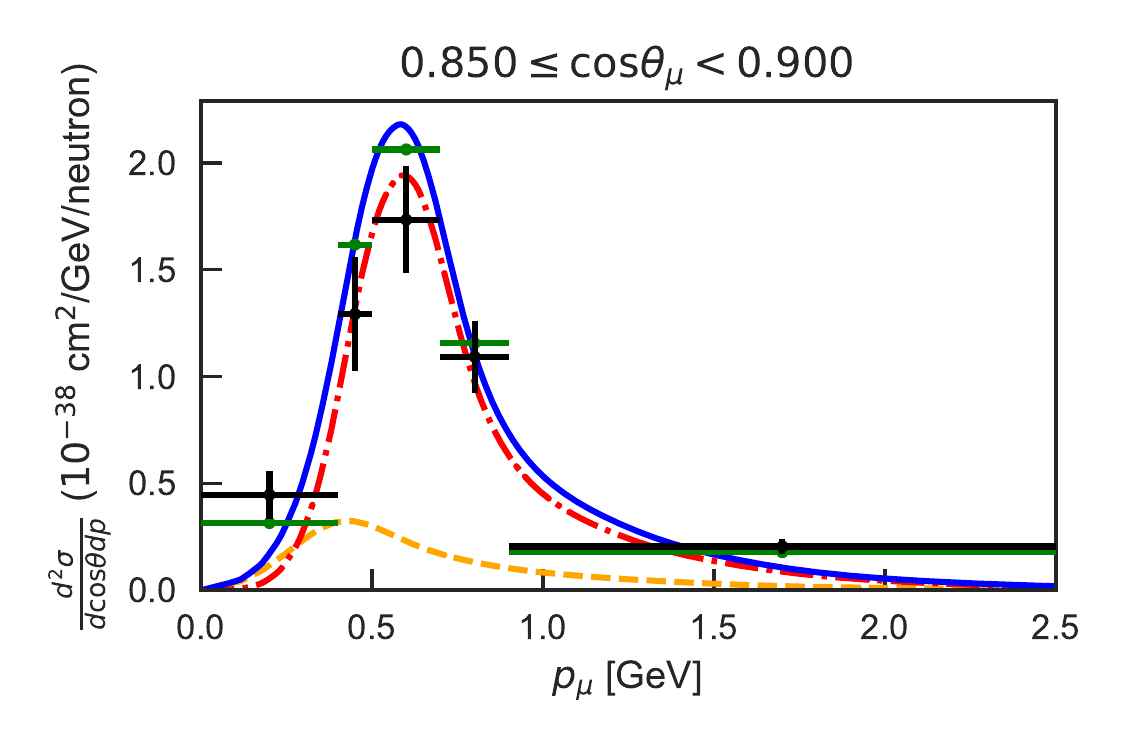}
  }
  \subfloat{
    \includegraphics[width=0.47\textwidth]{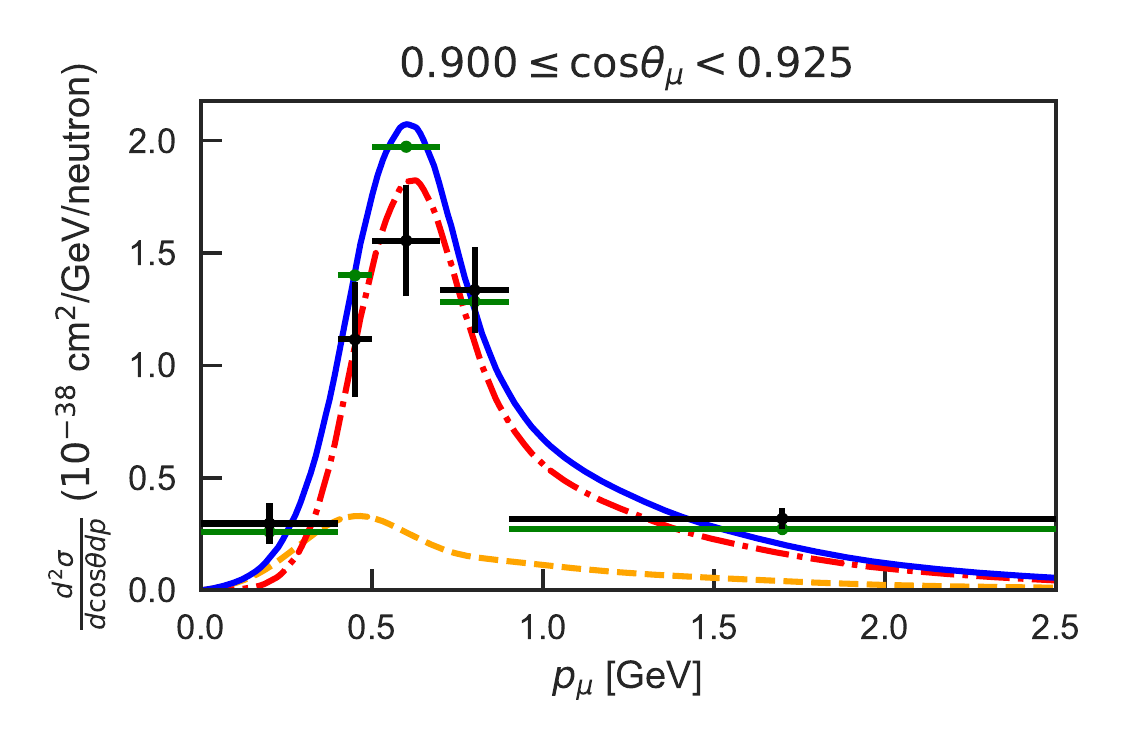}
  }\\
  \subfloat{
    \includegraphics[width=0.47\textwidth]{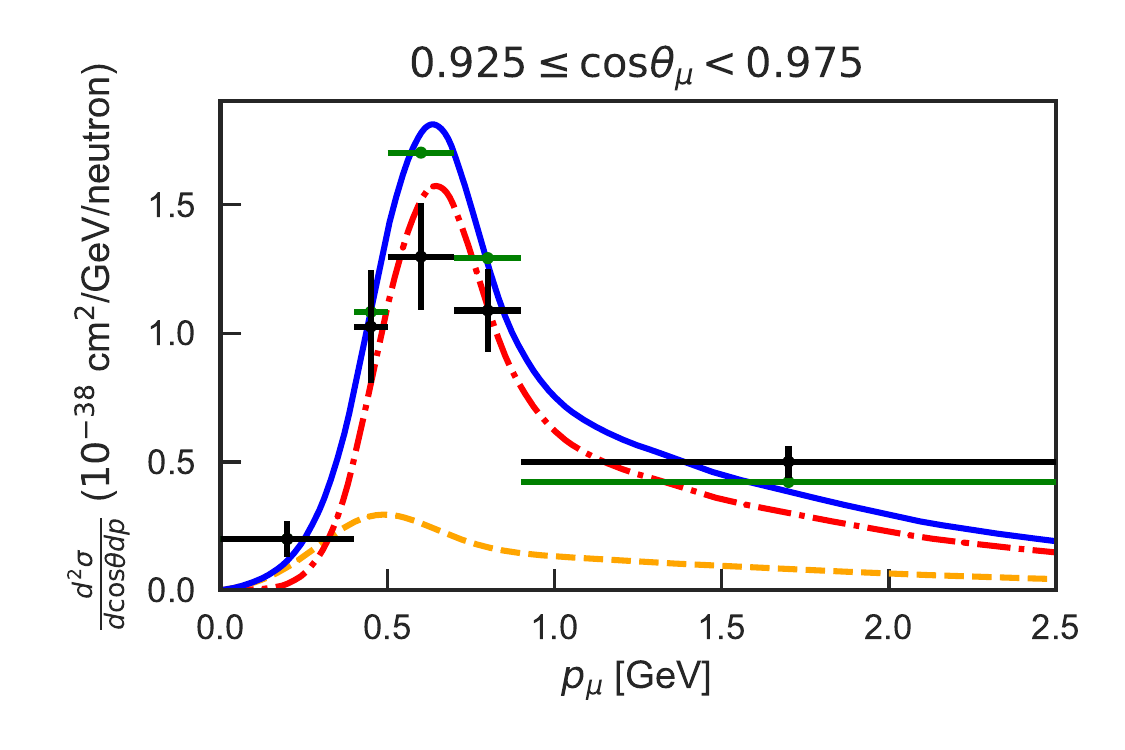}
  }
  \subfloat{
    \includegraphics[width=0.47\textwidth]{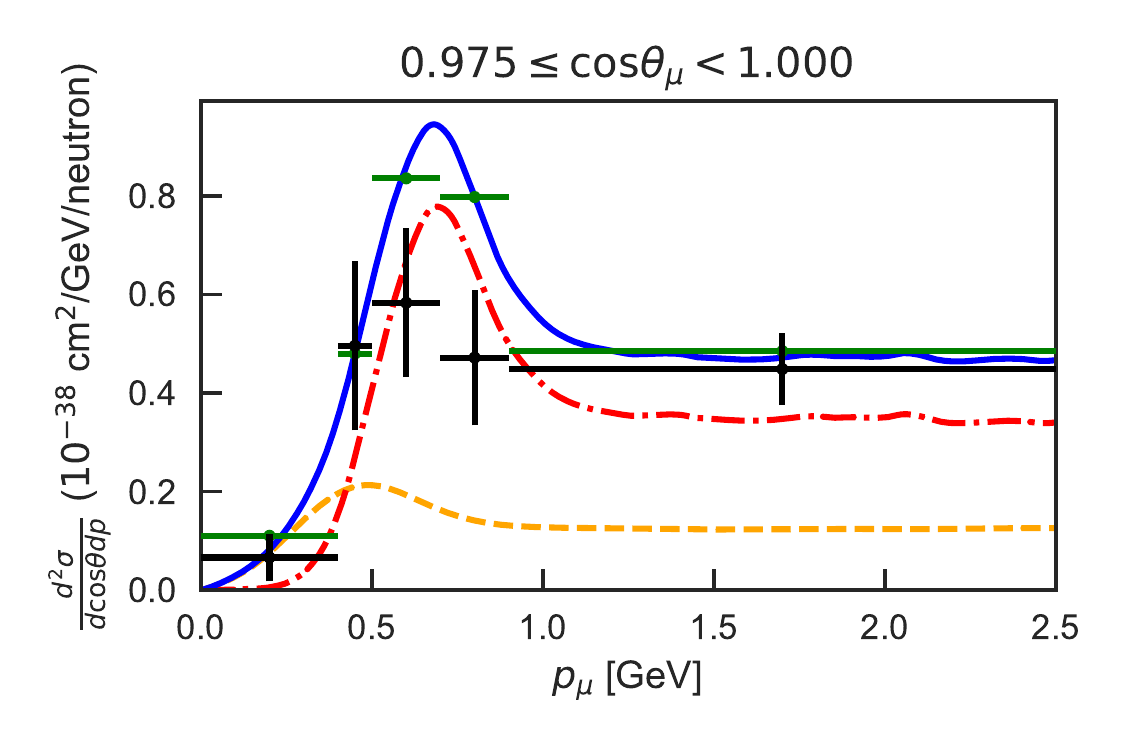}
  }\\
  \caption[Comparison against the SuSAv2 model]{A comparison of the
    \cczpi{} water cross section against the SuSAv2 model. Predictions
    for CCQE and 2p2h on oxygen are shown separately along with their
    sums and average over the bins.}
  \label{fig:compare_result_susa}
\end{figure*}

Finally, the result is compared to the Martini \emph{et
  al.}~\cite{martini_model} model predictions on carbon in
\cref{fig:compare_result_martini} and the SuSAv2
model predictions on water~\cite{superscaling, susav2} in \cref{fig:compare_result_susa}. Both models
are shown with and without 2p2h. Overall, our result agrees better with
an inclusion of 2p2h in most regions of phase space. This is
consistent with a similar comparison performed for the measurement on
\ce{C8H8} \cite{t2k_ccqe}.

The unfolding procedure allows events to migrate into and out of
regions in the phase space outside the binned area. Specifically, this
``out-of-range'' region refers to where $\cos \theta_\mu<0$ or $p_\mu
>\SI{5}{\GeV}$. As the selection efficiency here is very low, we
calculated a reduced, total cross section by integrating over only to
the explicitly binned areas of the $(p_\mu, \cos \theta_\mu)$ phase
space. The reduced, total cross section was found to be,
\begin{equation}
  \label{eq:reducedxsec}
  \begin{split}
   \sigma^{\mathrm{CC0\pi}}_{\numu{}\ce{H2O}}   = 0.95 \pm{} 0.08 (\mbox{stat}) \pm{} 0.06 (\mbox{det. syst.}) \\ 
   \SI[parse-numbers=false]{\pm{} 0.04 (\mbox{model syst.}) \pm{} 0.08 (\mbox{flux})  \times 10^{-38}}{\cm^2 \per n}.  
  \end{split}
   \end{equation}
This is significantly higher than the NEUT (GENIE) prediction of
\SI[parse-numbers=false]{0.66(0.68)\times 10^{-38}}{\cm^2 \per n}
primarily due to the disagreement between data and MC in the
high-angle regions, which cover a large portion of the reduced phase
space. The breakdown of the fractional uncertainties on
\cref{eq:reducedxsec} are given in \cref{table:fracuncert}.
\begin{table}[htb]
  \begin{ruledtabular}
  \begin{tabular}{lr}
    Source & Uncertainty [\%]\\
    \hline
    Statistics (data)  & \num{7.7}\\ 
    Statistics (MC) & \num{1.8}\\ 
    Detector mass & \num{1.5}\\ 
    Detector other & \num{6.3}\\ 
    FSI & \num{3.4}\\ 
    Cross Section & \num{3.2}\\
    Flux & \num{8.7}\\     
  \end{tabular}
  \end{ruledtabular}
  \caption{Fractional uncertainties on the total cross section. }
\label{table:fracuncert}
\end{table}

\section{\label{sec:conclusion}Conclusion}
Using the T2K near detector,
ND280, a CCQE-like, flux-integrated cross section on water was reported here
in the double-differential phase space of the outgoing muon.   Since the $\bar{\nu_{\mu}}$ contribution to this sample is insignificant, this is effectively a measurement on oxygen.  The
result complements previous T2K cross section measurements.   The fluxes used for this result, the extracted cross section values and covariance matrices for the errors are available at the following website: \url{http://t2k-experiment.org/wp-content/uploads/nd280data-numu-cc0pi-xs-on-h2o-2017.tar.gz}.

%The \pd{} subdetector of ND280 served as an interaction target with
%active scintillator and passive water modules~\cite{p0dnim}. Running
%the \pd{} in two different configurations, water-in and water-out,
%allowed us to perform a subtraction to obtain a distribution on the
%passive water layers. The selection is optimized for interactions with
%zero outgoing pions, and the raw distribution is corrected with a
%MC-based unfolding technique. 
A comparison to a double-differential
cross section result on carbon \cite{t2k_ccqe} shows good agreement
with the exception of a few low momentum bins in the high-angle region.  Comparisons were also performed to T2K simulations using
NEUT~\cite{neut} and GENIE~\cite{genie}.  Overall it appears that the tuned NEUT prediction is favored over GENIE, but in the angular regions of 
$0.7 < \cos{\theta_{\mu}} <0.85 $ it shows better agreement with GENIE.  Another
comparison to the Martini \emph{et al.}~\cite{martini_model} CCQE
prediction with and without 2p2h prefers the 2p2h
contribution.   A comparison to the SuSAv2
model predictions\cite{superscaling, susav2} also generally agrees with the data within errors, though the data points tend to be lower than the model in the more forward regions.

%While the current understanding of neutrino interactions on nuclei is
%expanding daily, there is still much to be explored. The importance of
%nuclear effects cannot be understated. Neutrino cross section
%measurements can help reduce such uncertainties. 
New techniques for CCQE measurements, such as exploring the transverse kinematics space,
or incorporating proton kinematics are ongoing in T2K. Future analyses from
T2K will have the benefit of higher statistics and can also include antineutrino cross sections.
% Perhaps, such measurements will
%disentangle some of the confusion surrounding our current models.

\section{Acknowledgements}

We thank the J-PARC staff for superb accelerator performance. We thank the
CERN NA61/SHINE Collaboration for providing valuable particle production data.
We acknowledge the support of MEXT, Japan;
NSERC (Grant No. SAPPJ-2014-00031), NRC and CFI, Canada;
CEA and CNRS/IN2P3, France;
DFG, Germany;
INFN, Italy;
National Science Centre (NCN) and Ministry of Science and Higher Education, Poland;
RSF, RFBR, and MES, Russia;
MINECO and ERDF funds, Spain;
SNSF and SERI, Switzerland;
STFC, UK; and
DOE, USA.
We also thank CERN for the UA1/NOMAD magnet,
DESY for the HERA-B magnet mover system,
NII for SINET4,
the WestGrid and SciNet consortia in Compute Canada,
and GridPP in the United Kingdom.
In addition, participation of individual researchers and institutions has been further
supported by funds from ERC (FP7), H2020 Grant No. RISE-GA644294-JENNIFER, EU;
JSPS, Japan;
Royal Society, UK;
the Alfred P. Sloan Foundation and the DOE Early Career program, USA.

\bibliographystyle{apsrev4-1}
\bibliography{CC0PiWater}% Produces the bibliography via BibTeX.

\end{document}